\documentclass[a4paper,11pt]{article}
\usepackage{pos}
\usepackage{float}
\usepackage{stackengine}
\usepackage[capitalise]{cleveref}
\usepackage[final]{pdfpages}

\title{Lattice Study of Spectator Effects in $b$-hadron Decays}

\author*[a]{Joshua Lin}
\author[a,b]{William Detmold}
\author[c]{Stefan Meinel}

\affiliation[a]{Center for Theoretical Physics,\\
  Massachusetts Institute of Technology, Cambridge, MA 02139, USA}

\affiliation[b]{The NSF AI Institute for Artificial Intelligence and Fundamental Interactions}

\affiliation[c]{Department of Physics,\\
  University of Arizona, Tucson, AZ 85721, USA}

\emailAdd{joshlin@mit.edu}

\abstract{The Heavy Quark Expansion (HQE) gives an expansion of inclusive decay rates of  $b$-hadrons as a simultaneous series in $\alpha_s$ and $1/m_b$, in terms of perturbatively defined coefficients and non-perturbative matrix elements. Spectator effects arise from the dimension-$6$ operators in the HQE in which along with the heavy quark, a light spectator quark from the hadron participates in the weak decay. In this work, we provide a lattice determination of the bare matrix elements that contribute to the spectator effects for the $B^+$-meson, $B_d$-meson and the $\Lambda_b$-baryon. The computations were performed on two separate lattice spacings of the RBC-UKQCD (2+1)-flavor Domain Wall Fermion ensembles. Renormalization and continuum extrapolation of the matrix elements have not yet been computed, and are left to future work. }

\FullConference{%
The 39th International Symposium on Lattice Field Theory,\\
8th-13th August, 2022,\\
Rheinische Friedrich-Wilhelms-Universität Bonn, Bonn, Germany
}

\begin{document}
\maketitle

\section{Introduction}
\begin{figure}
  \centering
\includegraphics[width=\textwidth]{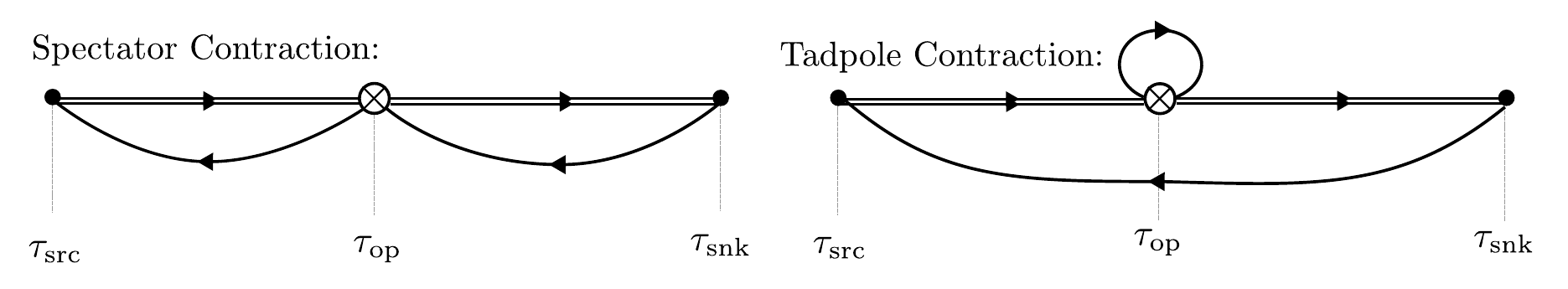}
\caption{Two types of contractions shown for a $B$-meson matrix element. The double-line corresponds to a heavy quark propagator, whereas the single lines are light quark propagators ($u,d$). The tadpole-like contraction shown on the right is not computed for either the meson or baryon in this work. }
\label{fig:contract}
\end{figure}

One of the theoretical approaches to studying the inclusive decays of hadrons including a $b$-quark is through the Heavy Quark Expansion (HQE), which expresses the hadron lifetime as a simultaneous $\alpha_s$ and $1/m_b$ expansion via an Operator Product Expansion (OPE) \cite{Lenz:2014jha,Shifman:2022spw}. The inclusive decay rate $\Gamma$ of a $b$-hadron $H_b$ is given by \cite{Lenz:2014jha}
\begin{equation}
  \begin{split}
  \Gamma (H_b) = \frac{G_F^2 m_b^5}{192 \pi^3} |V_{cb}|^2  \bigg\{ c_{3} \frac{\langle H_b | \overline{b} b|H_b \rangle}{2M_{H_b}} + &\frac{c_{5}}{m_b^2} \frac{\langle H_b | \overline{b} g_s \sigma_{\mu \nu} G^{\mu \nu} b | H_b \rangle}{2 M_{H_b}}  \\ +&\frac{c_{6}}{m_b^3} \sum_{i} \frac{\langle H_b | (\overline{b} \ \Gamma_L^{i} q) (\overline{q} \ \Gamma_R^{i} b) | H_b \rangle}{M_{H_b}} + O\left(\frac{1}{m_b^4}\right) \bigg\},
  \end{split}
\end{equation}
where the Fermi constant $G_F$ and the CKM-matrix element $V_{cb}$ appear after integrating out the $W$-boson. The factors of the mass of the hadron $M_{H_b}$ in the denominator reflect the relativistic normalisation of states. The decay rate is split into perturbatively calculable \cite{Altarelli:1991dx,Neubert:1996we} coefficients $c_n$, and non-perturbative QCD matrix elements of operators of increasing dimension, suppressed by inverse powers of the $b$-quark mass $m_b$. The $\Delta B = 0$, dimension $6$ QCD operators are written in terms of $\Gamma_L^i, \Gamma_R^i$, which are certain spin-colour matrices written explicitly in \cref{eq:explicit}. Even in the chiral limit, these QCD operators with different dimensions will mix with each other under renormalization due to the $m_b$ scale, so the $1/m_b$ series is not well-defined. It is convenient to match to Heavy Quark Effective Theory (HQET) to remove this scale and have a proper $1/m_b$ expansion, matching the QCD $b$-quark field onto a heavy (static) $Q$ field. Performing this expansion we obtain 
\begin{equation}
\begin{split}
\Gamma (H_Q)= \frac{G_F^2 m_b^5}{192 \pi^3} |V_{cb}|^2 \bigg\{ c_{3} \left[ 1 - \frac{\mu_{\pi}^2(H_Q) - \mu_G^2 (H_Q)}{2 m_b^2} + O\left(\frac{1}{m_b^3}\right) \right] + 2c_5 \left[ \frac{\mu_G^2(H_Q)}{m_b^2} + O \left( \frac{1}{m_b^3} \right) \right] \\ + \frac{c_6}{m_b^3} \sum_i \frac{\langle H_Q | (\overline{Q} \ \Gamma_L^i q) (\overline{q} \ \Gamma_R^i Q) | H_Q \rangle}{M_{H_b}} + O\left(\frac{1}{m_b^4} \right)\bigg\},
\end{split}
\end{equation}
where $\mu_\pi^2,\mu_G^2$ are the kinetic and chromomagnetic energies which are known precisely from spectroscopy \cite{Bigi:2011gf}, and we retain relativistic normalization of our states. Of the $O({1}/{m_b^3})$ contributions to the inclusive decay rate, the `spectator effects' in which light-quark degrees of freedom participate in the decay as well as the heavy quark dominate due to phase-space suppression of the other factors \cite{Neubert:1996we}. These come from the matrix elements of the dimension-6 operators
\begin{equation}
  \begin{gathered}
O^q_{V-A} = (\overline{Q} \gamma_\mu P_L q)(\overline{q} \gamma^\mu P_L Q), \quad 
O^q_{S-P} = (\overline{Q}  P_L q)(\overline{q}  P_R Q), \\
T^q_{V-A} = (\overline{Q} \gamma_\mu P_L T^a q)(\overline{q} \gamma^\mu P_L T^a Q), \quad  
T^q_{S-P} = (\overline{Q}  P_L T^a q)(\overline{q}  P_R T^a Q),
  \end{gathered}
  \label{eq:explicit}
\end{equation}
where $P_L = \frac{1-\gamma_5}{2},P_R=\frac{1+\gamma_5}{2}$ are the left/right projectors. Matrix elements of the above operators contain a contraction where a light valence quark from the hadron participates in the interaction, but also a tadpole-like contraction where the light quark in the operator is contracted in a loop as shown in \cref{fig:contract}. Only the former contraction is considered to be a spectator effect \cite{Neubert:1996we} as a light valence quark actually participates in the interaction. In this work we will only compute this contribution.

For the $B$-meson matrix elements, this tadpole subtraction is equivalent to considering isospin non-singlet versions of the operators. The contribution of the isospin-singlet piece vanishes when studying lifetime ratios such as $\tau(B^+)/\tau(B_d)$. Furthermore, isospin symmetry causes the operators to be protected from mixing with lower dimensional operators such as $\overline{b} b$. The matrix elements are conventionally parametrised for the mesonic states as \cite{Lenz:2014jha}

\begin{equation}
  \begin{split}
\frac{\langle B_d | (O^d_{V-A} - O^u_{V-A}) | B_d \rangle}{M_B} = f_B^2 B_1 M_B, &\quad \frac{\langle B_d | (O^d_{S-P} - O^u_{S-P}) | B_d \rangle}{M_B} = f_B^2 B_2 M_B, \\
\frac{\langle B_d | (T^d_{V-A} - T^u_{V-A}) | B_d \rangle}{M_B} = f_B^2 \epsilon_1 M_B, &\quad \frac{\langle B_d | (T^d_{S-P} - T^u_{S-P}) | B_d \rangle}{M_B} = f_B^2 \epsilon_2 M_B,
  \end{split}
  \label{eq:operators}
\end{equation}
where in the vacuum insertion approximation, $B_1=B_2 = 1, \epsilon_1=\epsilon_2 = 0$. By heavy-quark symmetry, the number of matrix elements is only $2$ and not $4$ for the $\Lambda_Q$ baryon, as the $(V-A)$ and $(S-P)$ type matrix elements are the same up to $1/m_b$ corrections (which would appear as $O(1/m_b^4)$ corrections to the lifetime, outside of the scope of this work). By isospin symmetry, the up-quark and down-quark type operators give the same matrix element for the $\Lambda_Q$ baryon. To isolate the spectator effects, we consider the normal-ordered version of the operator that does not have a tadpole-like contraction \cite{Neubert:1996we}. They are traditionally parametrised as \cite{Lenz:2014jha}
\begin{equation}
\frac{\langle \Lambda_Q | O^q_{V-A} | \Lambda_Q \rangle_\mathrm{tp. sub.}}{M_{\Lambda_b} } = f_B^2 L_1 M_B , \quad \frac{\langle \Lambda_Q | T^q_{V-A}  | \Lambda_Q \rangle_\mathrm{tp. sub.} }{M_{\Lambda_b}} = f_B^2 L_2 M_B,
\label{eq:bar_op}
\end{equation}
where the subscript reminds us that the matrix elements are tadpole subtracted. For the $\Lambda_Q$ baryon, the analogue of the vacuum-insertion approximation is the valence-quark approximation, which gives $L_2 = - \frac{2}{3}L_1$ \cite{Neubert:1996we}. 

Previous quenched lattice studies of the mesonic \cite{DiPierro:1998ty} and baryonic \cite{DiPierro:1999tb} matrix elements with Clover fermions have been performed. Preliminary results of an unquenched study were also presented \cite{Becirevic:2001fy}, with calculations of the $B$-meson matrix elements extrapolated to physical $M_{H_b}$. Our goal with this calculation is to perform the first lattice calculation of the quantities in \cref{eq:operators,eq:bar_op} with dynamical Ginsparg-Wilson (in particular, domain-wall) fermions, and improve on the precision and systematic control that was achieved in previous work. The fermion discretisation may be important as it is chiral symmetry that suppresses some mixing coefficients of the operators in \cref{eq:operators}.  Recently there have also been studies of alternative methods to calculating the inclusive decay rates using lattice methods but without an OPE \cite{Gambino:2022dvu}. A more precise calculation using the OPE would allow for better comparisons between the two methods, and could be crucial for further investigations into quark-hadron duality violations. Further work may also allow us to probe charm systems with the heavy-quark expansion, of interest due to recent experimental results \cite{LHCb:2021vll}.

\section{Lattice Setup and Fitting Procedure}

We use (2+1)-flavor Domain-Wall Fermion ensembles \cite{RBC:2014ntl} which were generated by the RBC-UKQCD collaboration using the Shamir DWF action \cite{Shamir:1993zy,Furman:1994ky}, with an Iwasaki gauge action. The parameters of the ensembles that are used are given in \cref{table:1}. 
\begin{table}[H]
  \centering
  \begin{tabular}{|l|l|l|l|}
  \hline
  Ensemble & Lattice Volume & $a^{-1}$ (GeV) & $m_\pi$ (MeV)  \\ \hline
  24I & $24^3 \times 64 (\times 16)$                   & $1.785(5)$                &         339.7(1.3)       \\ \hline
  32I & $32^3 \times 64 (\times 16)$                   & $2.383(9)$                &          302.4(1.2)        \\ \hline
  \end{tabular}
\caption{Parameters for the ensembles used in this study. Values taken from Table IX in \cite{RBC:2014ntl}.}
\label{table:1}
\end{table}

As we are working in lattice-HQET, the heavy quark is treated as a static Wilson line. To improve the statistical behaviour, we use Wilson Flow \cite{Luscher:2010iy}, which allows us to smear gauge links on the order of $\sqrt{8t}$, where $t$ is the flow time. To recover the continuum zero-flow-time matrix elements, it is enough to compute the matrix elements for a fixed lattice flow-time $a^{-2}t$ for various lattice spacings, and take the continuum limit. We compute correlation functions on flow time $a^{-2}t \in \{0.5,1.0,2.0\}$, as we found that at smaller flow-times the extraction of the matrix elements was unreliable. Domain-Wall-Fermion valence propagators have previously been computed for the light quarks for various source-sink separations, with Gaussian smearing in the spatial directions. The interpolating operators we will use to excite the $B$ meson and the $\Lambda_Q$ baryon are given by 
\begin{equation}
  B^{\dagger}_{t,s} = \overline{Q}_t \gamma^5 q_s,\quad {\Lambda}^{\dagger \alpha}_{t,s} = \epsilon_{abc} (\overline{u}^{a}_{s} (C \gamma^5) \overline{d}^{b T}_{s}) \overline{Q}^{c\alpha}_{t}, 
\end{equation}
respectively. The subscripts $t,s$ on the quark fields and the corresponding operators emphasize the fact that there is a choice of flow-time $t \in \mathbb{R}^+$ for the heavy-quark field, and also a choice of smearing $s \in \{\text{Local, Smeared}\}$ for the light quark. To extract the matrix elements, we first compute two and three point correlation functions using contraction code based on the Chroma library \cite{Edwards:2004sx} 
\begin{equation}
\begin{split}
C_2^{t s_1 s_2}(\tau_\mathrm{src},\tau_\mathrm{snk}) &= \langle \Omega | J_{t,s_1}(\tau_\text{snk}) J_{t,s_2}^\dagger(\tau_\text{src}) |\Omega \rangle, \\
 C_3^{t s_1 s_2}(\mathcal{O}; \tau_\mathrm{src}, \tau_\mathrm{op}, \tau_\mathrm{snk}) &= \langle \Omega | J_{t,s_1}(\tau_\text{snk}) \mathcal{O}_{t,L}(\tau_\text{op}) J_{t,s_2}^\dagger(\tau_\mathrm{src}) |\Omega \rangle,
\end{split}
\label{eq:corr}
\end{equation}
where $J^\dagger_{t,s} \in \{{B}^\dagger_{t,s}, {\Lambda}^{\dagger \alpha}_{t,s}\}$ is one of the hadronic creation operators. Since the heavy quark is static, all operators are at the same spatial location (whose value is suppressed in \cref{eq:corr}) and only differ in their temporal coordinate. The operator insertion $O$ is always local (unsmeared). To model the excited states, we use an $n$-state model 
\begin{equation}
  \begin{split}
  C_2^{t s_1 s_2}(\tau_\text{src},\tau_\text{snk}) &=  \sum_{i=1}^n Z^{i*}_{s_1} Z^i_{s_2} e^{-E_i (\tau_\text{snk}-\tau_\text{src})},\\
  C_3^{t s_1 s_2}(\mathcal{O}; \tau_\mathrm{src},\tau_\mathrm{op},\tau_\mathrm{snk}) &= \sum_{i,j=1}^n Z^{i*}_{s_1} Z^j_{s_2} e^{-E_i (\tau_\text{snk} - \tau_\text{op}) - E_j \tau_\text{op}}  \langle H_Q | \mathcal{O} | H_Q \rangle_\text{lat}, 
  \end{split}
\end{equation}
where the same excited states are assumed to be dominant in both the two and three-point fits. Combined fits are performed to the two- and three-point correlation functions, using the gvar and corrfitter packages \cite{Lepage:2001ym}. We perform fits over several different fitting ranges, indexed by two variables $2 \leq \tau^{(2)}_\mathrm{cut},\tau^{(3)}_\mathrm{cut} \leq 6$. This corresponds to fitting the $2$-point functions with minimum separation $\tau_\mathrm{snk} - \tau_\mathrm{src} \geq \tau^{(2)}_\text{cut}$, and the $3$-point functions with minimum source-operator and operator-sink separation $\tau_\mathrm{snk}-\tau_\mathrm{op},\tau_\mathrm{op}-\tau_\mathrm{src} \geq \tau^{(3)}_\text{cut}$. The two-point functions are also only fit with source-sink separation $\tau_{\mathrm{snk}}-\tau_\mathrm{src} \leq \frac{3T}{8}$ where $T$ is the total time-extent of the lattice, avoiding the need to model thermal effects. On the 24I ensembles the previously computed propagators allow for correlation functions to be determined for source-sink separations corresponding to $\tau_{snk}-\tau_\mathrm{src} \in \{2,\cdots,15\}$, whereas on the 32I ensemble there are fewer source-sink separations, $\tau_\mathrm{snk}-\tau_\mathrm{src} \in \{3,6,9,12,15\}$. The propagators have been computed for various pre-defined source patterns, so we first average the $C_2$ and $C_3$ functions over each gauge configuration for fixed separations, and then bootstrap over the configurations to estimate statistical uncertainties. 

Our analysis procedure is similar to the one performed in Ref. \cite{NPLQCD:2020ozd}. For a given fitting range indexed by $\tau_\mathrm{cut}^{(2)},\tau_\mathrm{cut}^{(3)}$, we use the Akaike Information Criterion \cite{1100705} to choose how many excited states to include in the fit. Explicitly, we increase the number of excited states to include in the fit, until the $\chi^2/N_\mathrm{d.o.f}$ improves by less than a fixed constant $\mathcal{A} = 0.1$. For all the fits $f$ that have $\chi^2/N_\mathrm{d.o.f} < 2$, associated to extracted matrix elements $M_f$ with statistical uncertainty $\delta M_f$, we weight them according to 
\begin{equation}
  \begin{split}
&w_f = \frac{p_f(\delta M_f)^{-2}}{\sum_{f'} p_{f'} (\delta M_{f'})^{-2}}, \quad \overline{M} = \sum_f w_f M_f, \\ \quad \delta_\mathrm{stat} \overline{M}^2 = \sum_f w_f \delta M_f^2, \quad &\delta_\mathrm{sys} \overline{M}^2 = \sum_f w_f (M_f - \overline{M_f})^2, \quad \delta \overline{M}^2 = \delta_\mathrm{stat} \overline{M_f}^2 + \delta_\mathrm{sys} \overline{M_f}^2,
  \end{split}
\end{equation}
where $p_f = \Gamma(N_\mathrm{d.o.f}/2,\chi^2/2)/\Gamma(N_\mathrm{d.o.f}/2)$ is the $p$-value of each respective fit, computed using the upper incomplete gamma function $\Gamma(\cdot,\cdot)$ and the standard gamma function $\Gamma(\cdot)$. $\overline{M} \pm \delta \overline{M}$ is the combined extracted matrix element, with combined statistical and systematic uncertainties. 

\section{Fit Results}

For display purposes, it is convenient to plot the ratio $R$ of the three-point to two-point functions as the $Z$-factors cancel, 
\begin{equation}
  R := \frac{C_3^{t s_1 s_2}(\mathcal{O}; \tau_\mathrm{src},\tau_\mathrm{op},\tau_\mathrm{snk})}{C_2^{t s_1 s_2}(\tau_\text{src},\tau_\text{snk})}, \quad \lim_{\tau_\mathrm{snk}-\tau_\mathrm{op},\tau_\mathrm{op}-\tau_\mathrm{src} \to \infty}R \to \langle H_b | \mathcal{O} | H_b \rangle_\text{lat}. 
\end{equation}
Note that we directly fit the correlation functions themselves rather than the ratios.

\begin{figure}
  \makebox[\textwidth][c]{\includegraphics[width=1.25\textwidth]{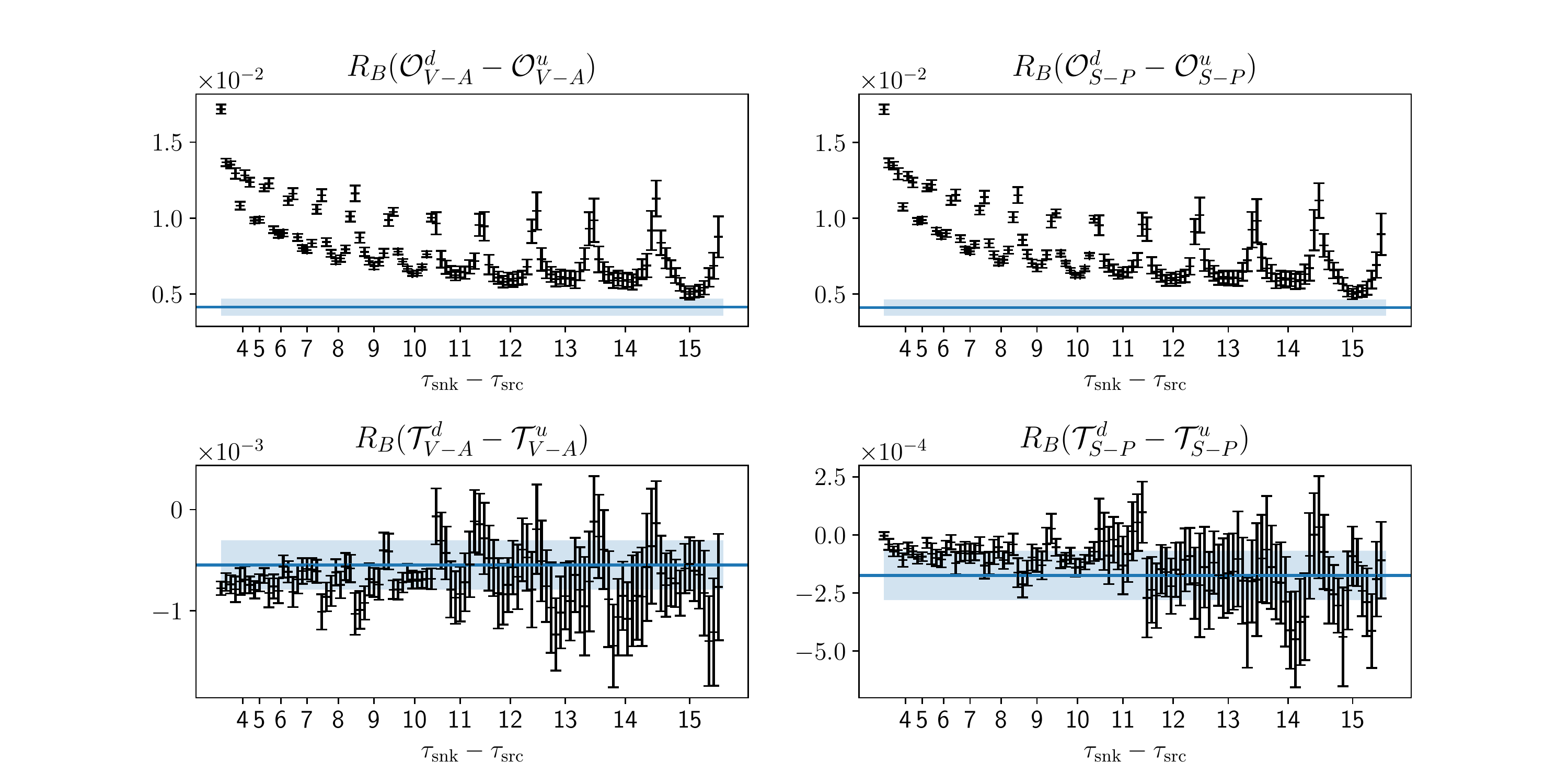}}
  \caption{Plots of the three-to-two point ratio functions for the $B$-meson matrix elements on the 24I ensemble, which should converge to the matrix element as $\tau_\mathrm{snk} - \tau_\mathrm{op},\tau_\mathrm{op}-\tau_\mathrm{src} \to \infty$. Within each choice of source-sink separation, there is an additional choice of the operator insertion time, and we can see plateaus forming when the operator is inserted far from both the source and sink. The blue line is the extracted matrix element, and the band represents the uncertainty (statistical and systematic). As expected, the color-mixed matrix elements in the bottom two panels are suppressed (here by about a factor of 10) compared to the unmixed ones. }
  \label{mes_fits}
\end{figure}

\begin{figure}
  \makebox[\textwidth][c]{\includegraphics[width=1\textwidth]{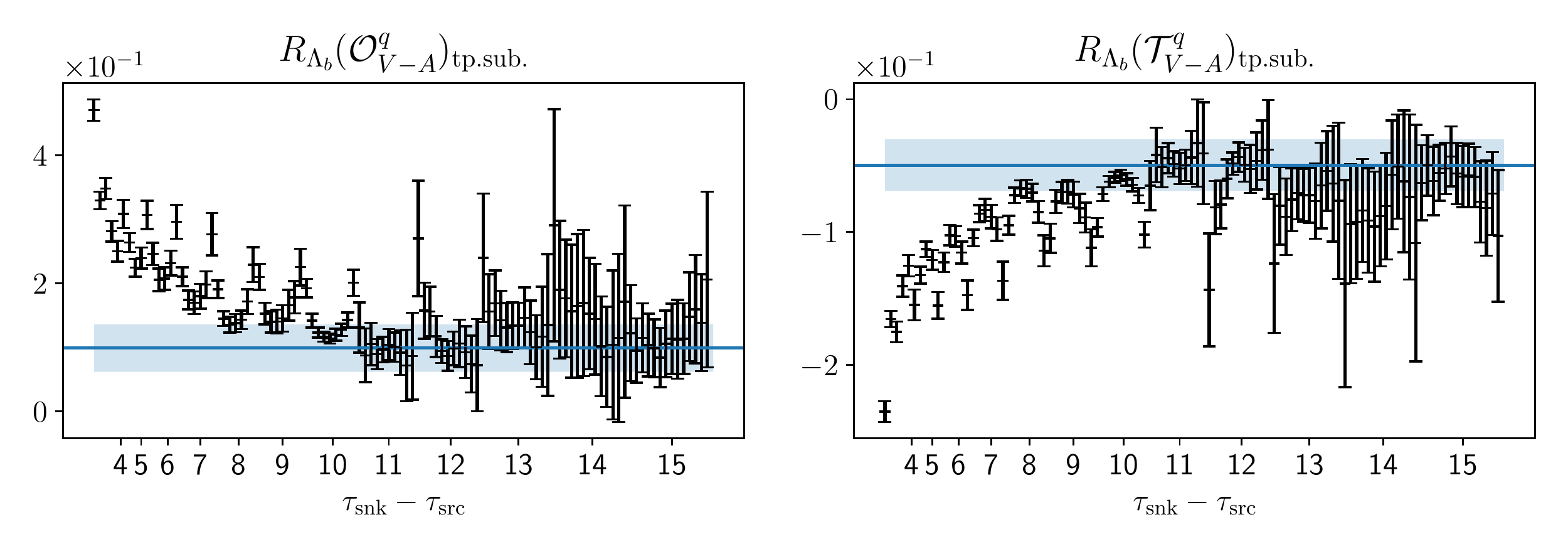}}
  \caption{Same plot as Fig \ref{mes_fits}, except for the two different baryon matrix elements on the 24I ensemble. There is no longer any suppression for the color-mixed operator, as the contraction pattern is different in the baryon case (There is no $\mathrm{Tr}(T^A)$ in the free limit).}
  \label{bar_fits}
\end{figure}

\begin{figure}
  \makebox[\textwidth][c]{\includegraphics[width=1.05\textwidth]{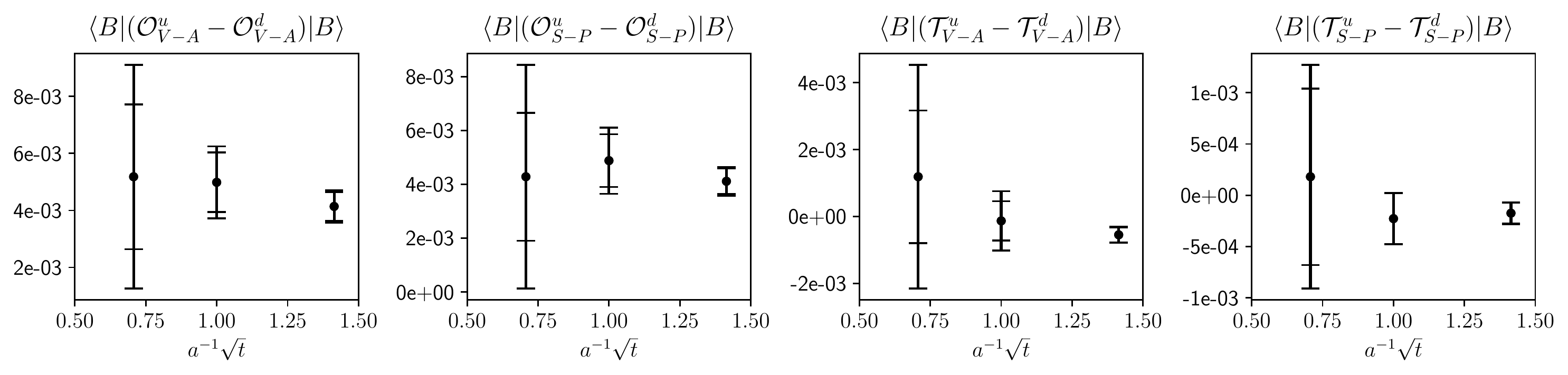}}
  \caption{Bare $B$-meson matrix elements as a function of the square root of the lattice flow time. The inner error-bar is the statistical uncertainty, and the outer error bar is the combined statistical and systematic uncertainty from the fits. Increasing flow time allows us to constrain the matrix elements much better. At large flow-time, the error is dominated by statistical uncertainty. }
  \label{mes_extrap}
\end{figure}

\begin{figure}
  \makebox[\textwidth][c]{\includegraphics[width=0.7\textwidth]{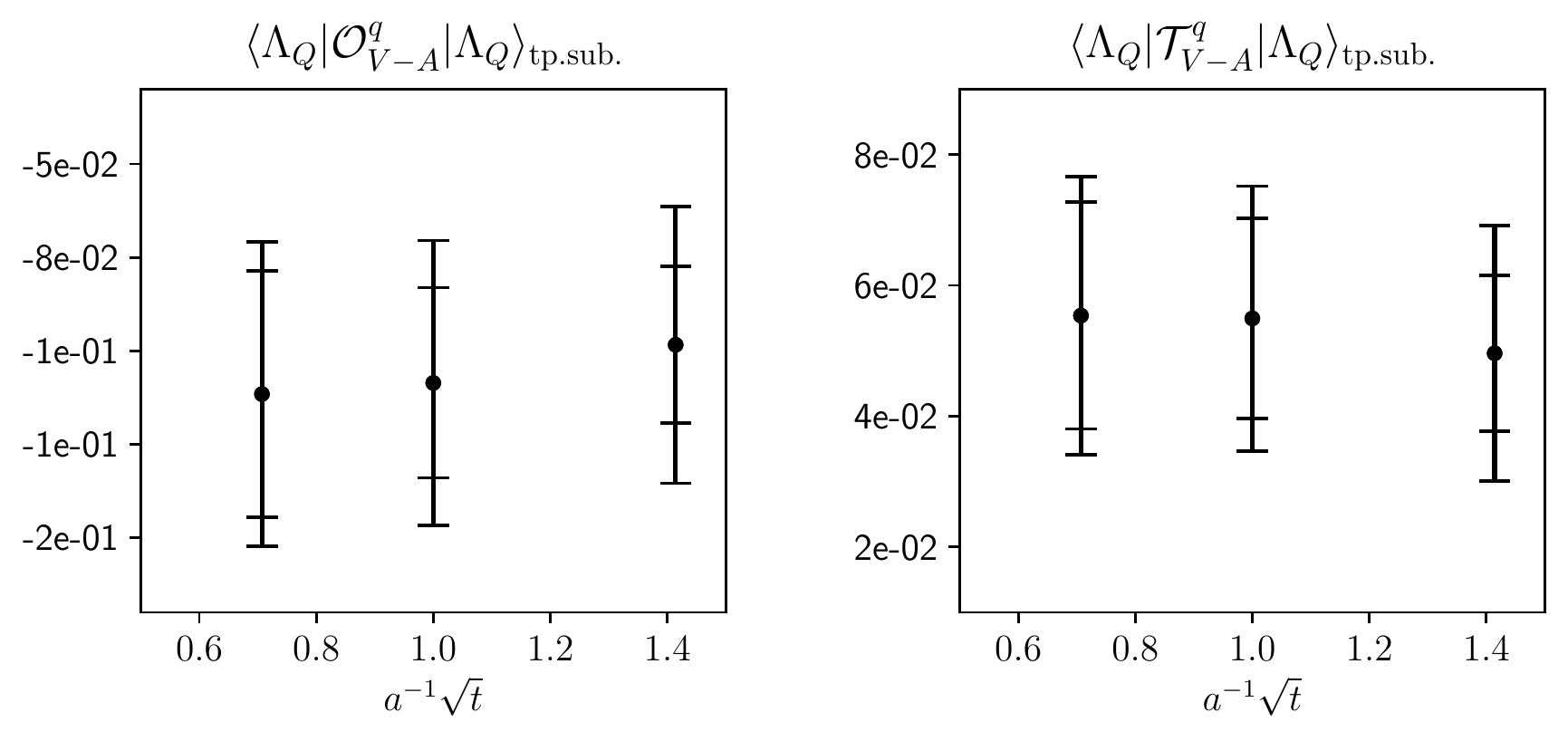}}
  \caption{Bare $\Lambda_Q$ matrix elements as a function of the square root of the lattice flow time. There is less variation in the size of the error-bar compared to the meson case, but the extraction still becomes unreliable for $a^{-1} \sqrt{t} < 0.5$. }
  \label{bar_extrap}
\end{figure}

The fits shown in Fig. \ref{mes_fits} and Fig. \ref{bar_fits} correspond to fits performed at the fixed lattice flow-time $a^{-2}t = 2.0$, for the meson and baryon matrix elements respectively. The ratios plotted are with both source and sink smeared. The errorbars on the datapoints were obtained by bootstrapping the ratio over the gauge configurations, and we can see clear plateaus forming in the $\tau_\mathrm{snk} - \tau_\mathrm{src} \to \infty$ limit for the non-color mixed meson matrix elements, and both the baryonic matrix elements. The horizontal band represents the total (stat+sys) uncertainty on the extracted matrix element. The flow-time dependence of the extracted mesonic matrix elements is shown in Fig. \ref{mes_extrap}, and for the baryonic matrix elements is shown in Fig. \ref{bar_extrap}.

\section{Conclusion and Outlook}

In this work we provide a new lattice determination of the dimension-$6$ matrix elements that contribute to spectator effects in the inclusive decay rates of $b$-hadrons. By computing these matrix elements at fixed lattice flow-times $a^{-2} t$ for various values of $a$, we can renormalize matrix elements on each ensemble, and taking the continuum limit we recover the zero flow-time, continuum matrix elements. We plan on performing a non-perturbative position-space renormalization, followed by a perturbative matching to $\overline{MS}$ in the continuum. We also plan on studying the tadpole-like contraction contributions to the baryon matrix elements, which have not yet been studied using lattice methods. 

\section*{Acknowledgements}
The computations for this work were carried out on facilities at the National Energy Research Scientific Computing Center, a DOE Office of Science User Facility supported by the Office of Science of the U.S. Department of Energy under Contract No. DE-AC02-05CH1123. SM is supported by the U.S.~Department of Energy, Office of Science, Office of High Energy Physics under Award Number DE-SC0009913. WD and JL are supported by the U.S. Department of Energy (DOE) grant DE-SC0011090. WD is also supported in part by the SciDAC5 award DE-SC0023116 and by the National Science Foundation under Cooperative Agreement PHY-2019786 (The NSF AI Institute for Artificial Intelligence and Fundamental Interactions, http://iaifi.org/).

\bibliographystyle{utphys-noitalics}
\bibliography{refs_new}


\end{document}